\documentclass[journal=jacsat,manuscript=article]{achemso}
\usepackage[version=3]{mhchem} 
\usepackage{graphicx}
\usepackage{xcolor}
\usepackage{amsmath,amssymb}
\usepackage{upgreek}
\usepackage{verbatim}
\usepackage{comment}

\title[Silicon Nanowires]{Quantum Interference in Silicon 1D Quasi-Ballistic Junctionless Nanowire Field Effect Transistors}

\author{Felix J. Schupp}
\affiliation[University of Oxford, Department of Materials, 16 Parks Road, Oxford OX1 3PH, UK]{University of Oxford, Department of Materials, 16 Parks Road, Oxford OX1 3PH, UK\\}

\author{Muhammad M. Mirza}
\affiliation[University of Glasgow, School of Engineering, Rankine Building, Oakfield Avenue, Glasgow, G12 8LT, United Kingdom]{University of Glasgow, School of Engineering, Rankine Building, Oakfield Avenue, Glasgow, G12 8LT, United Kingdom\\}

\author{Donald A. MacLaren}
\affiliation[University of Glasgow, SUPA School of Physics and Astronomy, Kelvin Building, University Avenue, Glasgow, G12 8QQ, United Kingdom]{University of Glasgow, SUPA School of Physics and Astronomy, Kelvin Building, University Avenue, Glasgow, G12 8QQ, United Kingdom\\}

\author{G. Andrew D. Briggs}
\affiliation[University of Oxford, Department of Materials, 16 Parks Road, Oxford OX1 3PH, UK]{University of Oxford, Department of Materials, 16 Parks Road, Oxford OX1 3PH, UK\\}

\author{Douglas J. Paul}
\affiliation[University of Glasgow, School of Engineering, Rankine Building, Oakfield Avenue, Glasgow, G12 8LT, United Kingdom]{University of Glasgow, School of Engineering, Rankine Building, Oakfield Avenue, Glasgow, G12 8LT, United Kingdom\\}

\author{Jan A. Mol}
\affiliation[University of Oxford, Department of Materials, 16 Parks Road, Oxford OX1 3PH, UK]{University of Oxford, Department of Materials, 16 Parks Road, Oxford OX1 3PH, UK\\}
\email{Jan.Mol@materials.ox.ac.uk}

\begin{document}
\begin{abstract}
We investigate the low temperature transport in 8 nm diameter Si junctionless nanowire field effect transistors fabricated by top down techniques with a wrap-around gate and two different activated doping densities.  First we extract the intrinsic gate capacitance of the device geometry from a device that shows Coulomb blockade at 13 mK with over 500 Coulomb peaks across a gate voltage range of 6 V indicating the formation of a single island in the entire nanowire channel. In two other devices, doped Si:P $4\times10^{19}\,\text{cm}^{-3}$ and $2\times10^{20}\,\text{cm}^{-3}$, we observe quantum interference and use the extracted gate coupling to determine the dominant energy scale and the corresponding mean-free paths. For the higher doped device the analysis yields a mean free path of $4\pm2\,\text{nm}$, which is on the order of the average dopant spacing and suggests scattering on unactivated or activated dopants. For the device with an activated dopant density of $4\times10^{19}\,\text{cm}^{-3}$ the quantum interference effects suggest a mean free path of $10\pm2\,\text{nm}$, which is comparable to the nanowire width, and thus quasi-ballistic transport. A temperature dependent analysis of Universal Conductance Fluctuations suggests a coherence length above the nanowire length for temperatures below 1.9 K and decoherence from 1D electron-electron interactions for higher temperatures. The mobility is limited by scattering on impurities rather than the expected surface roughness scattering for nanowires with diameters larger or comparable to the Fermi wavelength. Our measurements therefore provide insight into the performance limitations from dominant scattering and dephasing mechanisms in technologically relevant silicon device geometries.
\end{abstract}

\maketitle

Silicon nanowires have been extensively studied with diameters down to below 5 nm\cite{Zhong:2005a,Yi:2011} and for a wide range of applications including electronics\cite{Nanowire:2008,Colinge:2010fk}, qubits\cite{SpinQubitNanowire}, biosensors\cite{CancerNanowire, Zheng:2010}, colour selective photodetectors\cite{Park:2014}, photovoltaics\cite{PVnanowire} and thermoelectric generators\cite{Thermoelectric}. Short channel effects and poor electrostatic control of the channel in two-dimensional transistors such as MOSFETs have led to significant work on nanowire transistors where a wrap around or Omega gate provides strong electrostatic control of the channel\cite{Kuhn:2012} and in the smaller nanowires can lead to one-dimensional (1D) electron transport\cite{Mirza:2017}.
1D transport has been studied extensively in carbon-nanotubes, metal and semiconductor nanowires\cite{Cao:2005,VanWeperen:2013,Wu:2004}. Strong radial confinement in these systems leads to the formation of subbands, that can be populated or depleted with excellent electrostatic control in e.g. multi-gate geometries. In ultra clean devices, with scattering lengths longer than the one-dimensional transport channel, quantum interference leads to Fabry-Perot type transport where energy is only dissipated at the source and drain contacts\cite{Dirnaichner:2016}. If the dominant scattering processes could be identified and controlled in technologically relevant CMOS devices, this limit could be reached and lead to the development of low power transistors. The short channels required for such devices are challenging to realize with different doped regions, such that a homogeneously doped ''junctionless" design is a promising candidate.\cite{Colinge:2010fk} Here we demonstrate gate all-around junctionless silicon nanowires, that were fabricated from silicon-on-insulator wafers and reach the 1D quasi-ballistic limit with a mean free path larger than the diameter for a doping density of Si:P $4\times10^{19}\,\text{cm}^{-3}$. The dominant scattering process is determined to be neutral impurity scattering rather than surface roughness scattering. This is the result of high doping densities, highly optimized fabrication and low interface trapped charge density.\\ 

\begin{figure}
\includegraphics[width=3.33in]{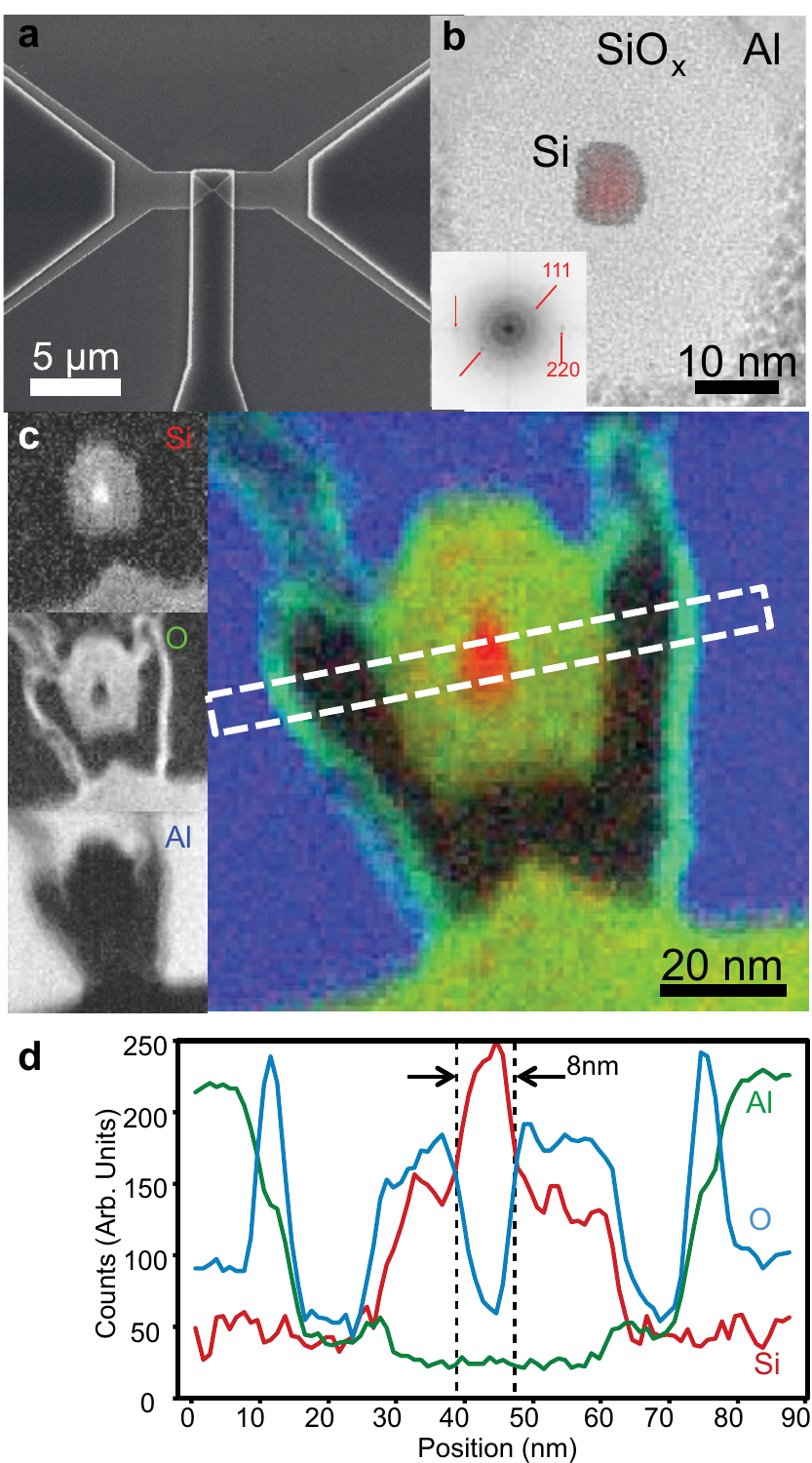}
\caption{a) Top view of the device taken using a SEM. b) Cross-sectional TEM image of a nominal 10 nm diameter nanowire with the lattice fringes picked out in red using a filtered Fourier transform  (inset) to select lattice fringes. c) Elemental mapping of the same nanowire. The main image is a composite of Si (red), Oxygen (green) and Aluminium (blue) signals with the individual maps inset to the left. d) Integrated signal intensity profiles across the dashed region of the RGB map used to determine the wire width. \label{Device}}
\end{figure}

We have investigated P doped Si nanowires  with a length of $L=150$~nm and a wrap-around aluminium gate that surrounds the entire nanowire (for fabrication details see Supplementary Information). The physical parameters of the silicon nanowire, such as diameter, length, crystallinity, and interface quality are crucial in determining its transport properties. We therefore for characterize the fabrication process via electron energy loss spectroscopy scanning transmission electron micrograph (EELS-STEM) and capacitance-voltage ($C$-$V$) measurements. Fig.~\ref{Device}a is an EELS-STEM image of the cross-section of a nanowire transistor with an inner diameter of $8\pm0.5$~nm, as determined from the transition in relative Si and O concentrations. The EELS-STEM data shows that the gate is not perfectly wrapped around the nanowire resulting in a vacuum-gap underneath a section of the gate oxide (see Fig. \ref{Device}a). The previously reported on-current to off-current ratio above $10^8$ with subthreshold slope of $66\,\text{mV/dec}$ at $300\,\text{K}$ demonstrates that this gap does not significantly affect the electrostatic control of the channel by the gate\cite{Mirza:2017,Georgiev:2017}. We will confirm the effectiveness of the wrap-around gate at $13\,\text{mK}$ by analyzing the gate capacitance based on the electron addition energy below. A high resolution STEM image in Fig. \ref{Device}b with the [111] and [220] lattice fringes highlighted in red confirms that the Si is a single crystal lattice orientated along [110] and surrounded by the amorphous SiO$_2$. Fig. \ref{Device}c shows a top view scanning electron microscope (SEM) image of the same device structure. Although the gate covers the channel as well as a part of the source and drain contacts, the transport is dominated only by the nanowire-channel as the contacts are degenerately doped and therefore metallic\cite{Georgiev:2017,Mirza:2017}.\\

\begin{figure}
\includegraphics[width=3.33in]{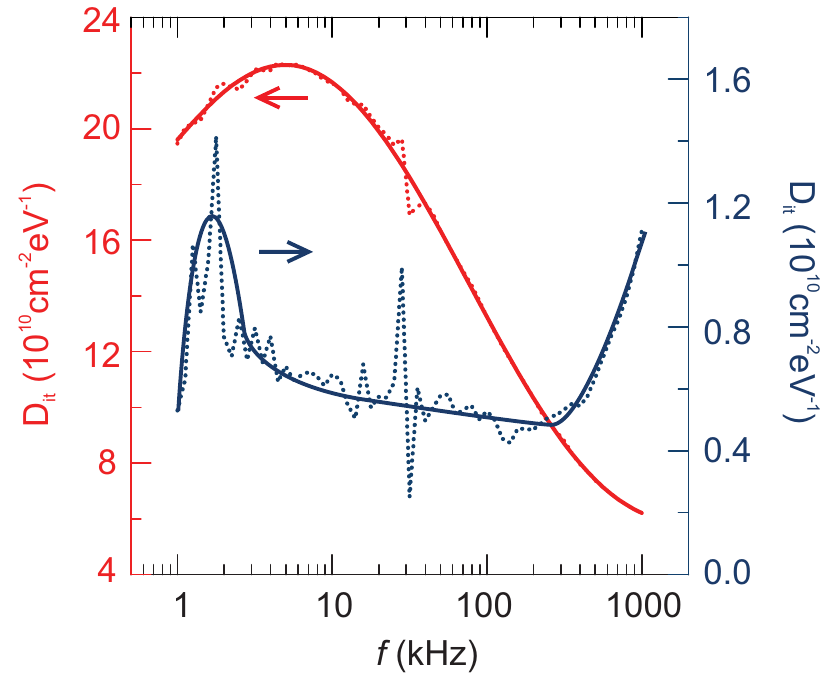}
\caption{Density of interface traps $D_\text{it}$ as a function of frequency $f$ before the forming gas anneal (red) and after forming gas anneal (blue). \label{Dit}}
\end{figure}

In addition to the crystal quality of the silicon, the nanowire performance is determined by the quality of the surface passivation. The room temperature performance and temperature scaling behavior of the nanowires above 10 K has been published elsewhere\cite{Georgiev:2017,Mirza:2017,Busche:2014fk}. To investigate the role of deep interface trap states we examine the $C$-$V$ characteristics of 100 $\upmu$m circular MOS capacitors fabricated on (100) crystal oriented n-type silicon substrate (N$_D$=3.5 $\times$ $10^{−15}\,\text{cm}^{-3}$) with 10 nm thermally grown SiO$_2$ that were processed in the same oxidation furnace in which the nanowires were produced. Figure~\ref{Dit} shows the presence of mid-gap states in the thermally grown oxide with a large interface trap density $D_{\text{it}}$ = 2.3 $\times$ $10^{11}$ cm$^{-2}$~eV$^{-1}$ measured at 1~MHz (red). Forming gas passivates the dangling bonds and trap charges with hydrogen atoms and lowered $D_{\text{it}}$ by over an order of magnitude down to 1.3 $\times$ $10^{10}$ cm$^{-2}$~eV$^{-1}$ (blue). From the $C$-$V$ measurements, we conclude that there will be on average only one deep interface trap on the surface of each nanowire. Therefore, scattering of surface states will be negligible.\\
In the following we characterize the transport through our devices using two types of measurements. In both measurement types the gate and bias voltage is supplied by battery powered DC measurement electronics. For the temperature dependence in Fig. 3 we use a lock-in technique at a frequency of 1337.3 Hz with an excitation amplitude of 50 $\upmu$V for temperatures below 1.45 K and 100 $\upmu$V for higher temperatures. All other measurements were carried out in DC using a current amplifier that is also powered by a battery. The measurement system was carefully filtered to ensure low electron temperatures and the associated series resistance was calibrated out of in the data. With exception of the conductance in Fig. \ref{Tdep} all derivatives were calculated numerically.\\

\begin{figure}[ht!]
\includegraphics[width=3.33in]{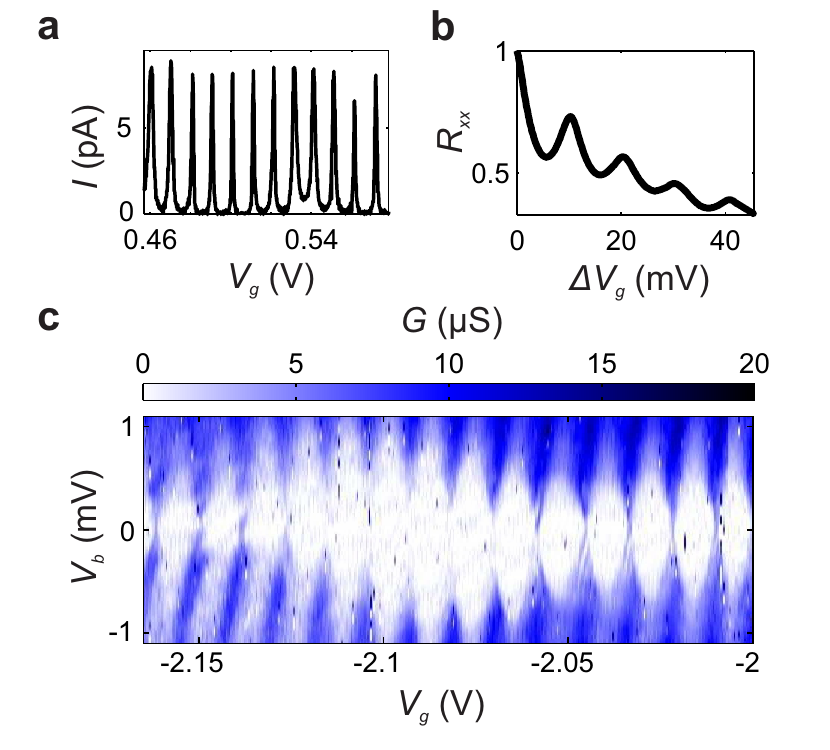}
\caption{Data from device A: a) Current $I$ at $0.1\,\text{mV}$ bias as a function of gate voltage $V_\text{g}$ through a Si nanowire device with diameter $8\pm0.5\,\text{nm}$, length $L=150\,\text{nm}$ and P doping of $2\times10^{-20}\,\text{cm}^{-3}$. b) Normalized autocorrelation function $R_\text{xx}$ of the current in a larger gate voltage window from -4.5 V to +2 V. c) Conductance $G$ as a function of gate voltage $V_\text{g}$ and bias voltage $V_\text{b}$.\label{CB}}
\end{figure}

\begin{figure}
\includegraphics[width=3.33in]{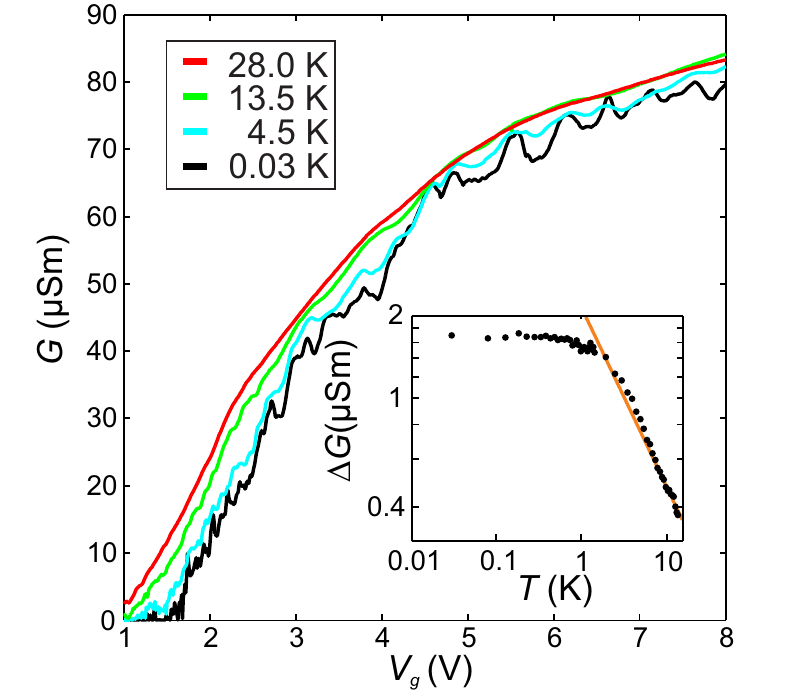}
\caption{Temperature dependent conductance measured for device B: a) Conductance $G$ as a function of gate voltage $V_\text{g}$ at four different temperatures from 30 mK to 28 K (Temperature traces at 82 temperatures are shown in the Supplementary Information). Inset: Root-mean square of the conductance traces $\Delta G$ after subtracting a fourth degree polynomial fit to isolate the fluctuations from the background as a function of temperature $T$. The brown line is a fit with $\Delta G\propto T^{-\gamma}$ resulting in $\gamma=0.67\pm0.04$.\label{Tdep}}
\end{figure}

Fig. \ref{CB} presents data from device A -- a $8\pm0.5\,\text{nm}$ diameter nanowire with a $150\,\text{nm}$ channel and an activated P doping density of $2\times10^{20}\,\text{cm}^{-3}$. The activated carrier densities were obtained from temperature dependent Hall measurements on large micrometer scale Hallbar devices between 1 and 300 K with geometrical uncertainties in the carrier density at 1 K below 1\%\cite{Mirza:2014}. The 1 K Hall density is quoted as the activated density to remove any thermally activated trapped states from the density. The device shows signatures of single electron transport at $13\,\text{mK}$ with more than 500 evenly spaced Coulomb peaks over a large gate voltage range from -4.5 to +2 V. Fig. \ref{CB}a shows a small fraction of these features in current and the full gate-voltage range is presented in the Supplementary Information.\\
To characterize the peak-spacing over the entire gate range, we calculate the autocorrelation function $R_{xx} = \int{I(V_{g})I(V_{g}-\Delta V_{g})\text{d}V_{g}}$ of the current as a function of gate voltage in Fig. \ref{CB}b. The autocorrelation demonstrates oscillations that correspond to a spacing of the Coulomb blockade peaks $\Delta V_\text{CB}=10\pm 1\, \text{mV}$. In Fig. \ref{CB}c we show the corresponding Coulomb diamonds in conductance $G$ as a function of gate voltage $V_\text{g}$ and bias voltage $V_\text{b}$.\\
The regular spacing of the Coulomb blockade peaks indicates that the charge island has a fixed capacitance, and therefore a fixed size, which implies that a single island forms within the channel that is defined by two tunnel barriers and persists over a large gate range. We estimate the island length by comparing the capacitance to the gate $C_\text{g}$ taken from the data to a theoretical value that follows from a simple cylindrical capacitor model (see Supplementary Information). The model for a 150 nm nanowire yields $C_\text{g}=15.2$ aF, while the capacitance from the data in Fig. \ref{CB} is extracted from the spacing on the gate voltage axis at zero bias $\Delta V_\text{CB}=10\pm 1\, \text{mV}$\cite{VanHouten:1992}:\\
\begin{equation}
C_\text{g}=e/\Delta V_\text{CB}=16\pm 2\,\text{aF}
\end{equation}
where $e$ is the elementary charge.\\
From the good agreement of the experimental and theoretical values for the capacitance, we conclude that the tunnel barriers are located close to the ends of the nanowire channel, i.e. the charge island has the same length as the nanowire.
The formation of tunnel barriers at the ends of the channel are either related to imperfect lithography optimization due to proximity effects at the ends of the nanowire, strain and/or the accumulation of impurities.\cite{Shin:2010} The extracted value of the capacitance is intrinsic to the nanowire geometry and therefore valid for all geometrically identical devices.\\

\begin{figure}
\includegraphics[width=3.33in]{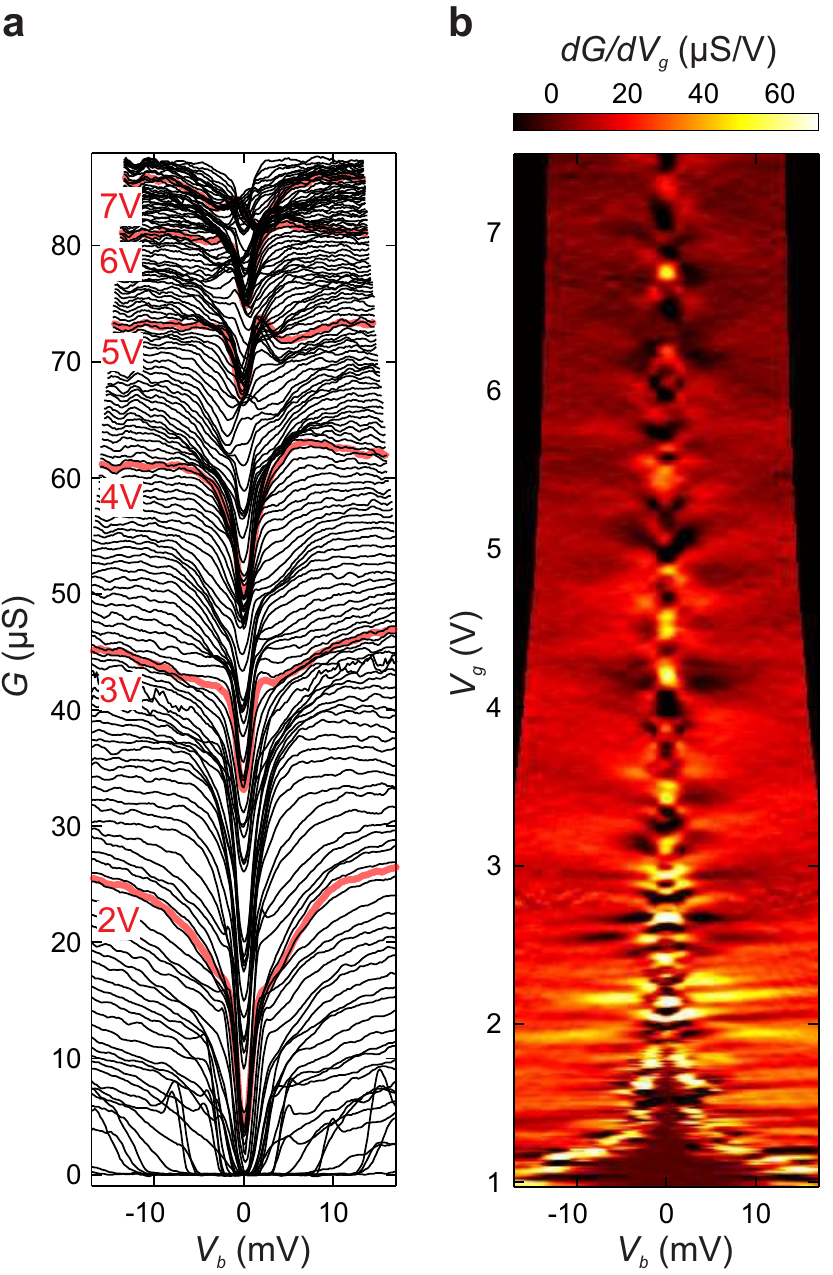}
\caption{Conductance fluctuations measured from device B (corrected for a series resistance of 12.5 k$\Omega$): a) Conductance traces as a function of bias $V_\text{b}$ at gate voltages from $0.97\,\text{V}$ to $7.5\,\text{V}$ in a silicon nanowire device with length 150 nm, diameter $8\pm 0.5$ nm and activated P doping of  $4\times10^{19}\,\text{cm}^{-3}$. The red lines indicate the traces at 2, 3, 4, 5, 6, and 7 V gate voltage. b) Change in conductance $dG/dV_\text{g}$ as a function of bias voltage $V_\text{b}$ and gate voltage $V_\text{g}$ in the same device.\label{waterfallandstab}}
\end{figure}

\begin{figure}
\includegraphics[width=3.33in]{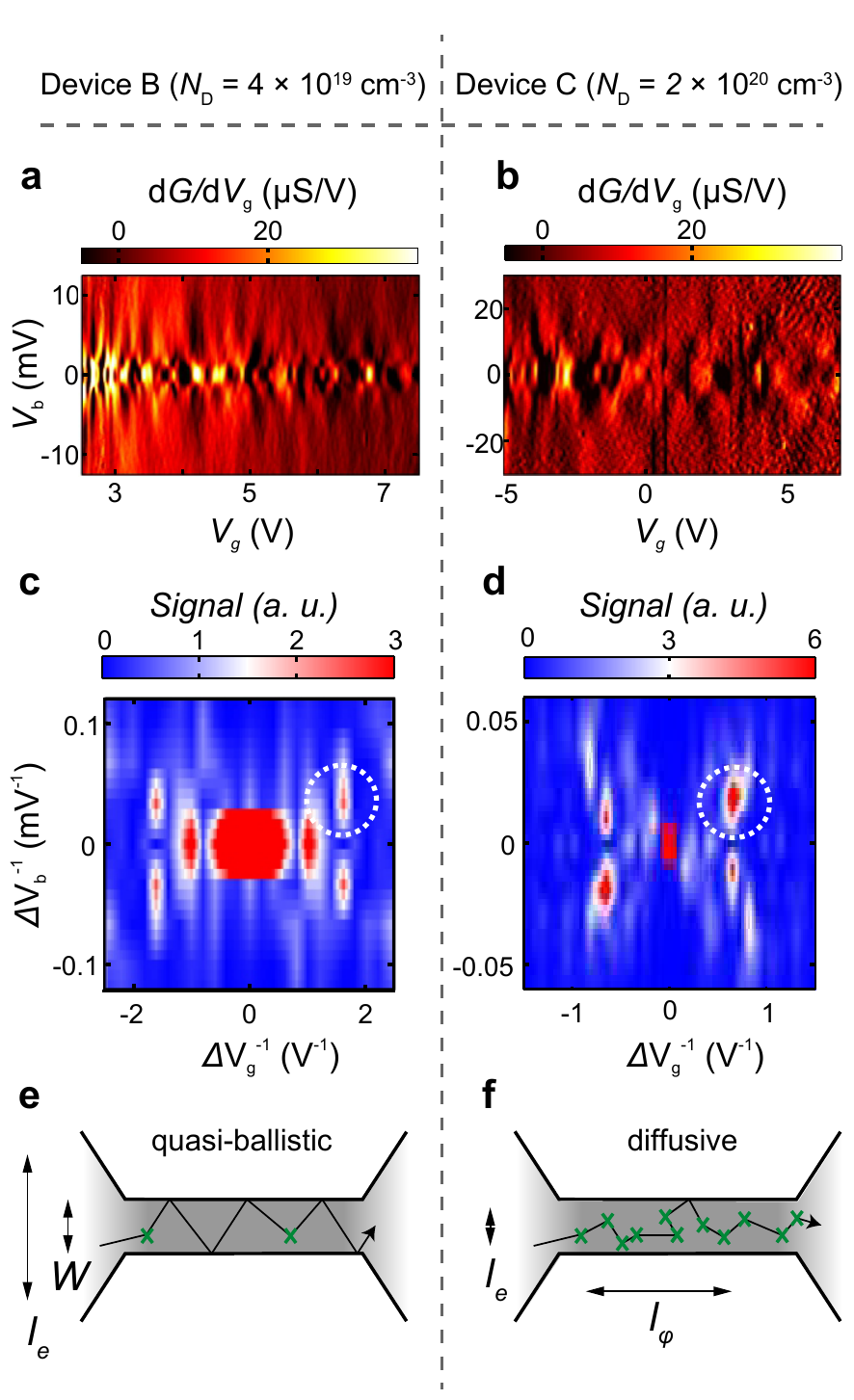}
\caption{Data from device B with diameter 8 nm and activated P doping of $4\times10^{19}\,\text{cm}^{-3}$ at 13 mK: a) Transconductance $dG/dV_\text{g}$ as a function of gate voltage $V_\text{g}$ and bias voltage $V_\text{b}$, c) Fourier transform of the autocorrelation of the data in a) with one of four symmetric hotspot circled in white. Data from device C with diameter 8 nm and activated P doping of $2\times10^{20}\,\text{cm}^{-3}$ at 4 K: b) Transconductance $dG/dV_\text{g}$ as a function of gate voltage $V_\text{g}$ and bias voltage $V_\text{b}$, d) Fourier transform of the autocorrelation from c) with a hotspots circled in white. e) Schematic depiction of a typical electron path through a nanowire with diameter $W$ and length $L$ in the quasi ballistic transport regime. The scattering events, denoted by the green markers, allow for coherent reflections in the radial direction. f) Schematic depiction of a typical electron path through a nanowire with diameter $W$ and length $L$ in the diffusive transport regime. Scattering, denoted by the green markers, is so frequent that coherent modes in radial direction cannot form.\label{FFTauto}}
\end{figure}

Transport data from measurements on device B are shown in Fig. \ref{Tdep}, \ref{waterfallandstab} and \ref{FFTauto}a,b. Device A and B are geometrically identical, with 150 nm channel length and $8\pm0.5\,\text{nm}$ diameter, but device B has a lower activated doping concentration of $4\times10^{19}\,\text{cm}^{-3}$. The low temperature conductance data do not show the regular Coulomb blockade pattern over a large gate range that we observed in device A. The associated Coulomb diamonds also strongly vary in size (see Supplementary Information) and often do not close completely, which is a signature of transport through one or more islands in the nanowire with varying sizes as a function of gate voltage. It is therefore likely that the charge islands form as a result of potential variations along the nanowire and not solely due to potential barriers at the ends of the nanowire.\\
While transport in device A is dominated by Coulomb blockade over the entire measured gate voltage window, device B is characterized by a Coulomb blockade region followed by increasing conductance without blockade (see Supplementary Information) at higher gate voltages ($V_{g}>1.7$~V). Above this threshold voltage we observe fluctuations in the conductance that we attribute to quantum interference effects that are analyzed in the next sections.\\ 
In nanoelectronics quantum interference is the interference of partial charge carrier waves e.g. the counter-propagating partial waves between two reflecting points. The required coherent scattering can occur on any stationary boundary such as the ends of the nanowire (longitudinal Fabry-P\'erot type interference)\cite{Kretinin:2010,Gehring:2016}, random potential fluctuations in the nanowire from e.g. impurities (Universal Conductance Fluctuations)\cite{Cahay:1988,Elm:2015} or transverse modes due to the confinement in the cross section of the nanowire (subbands or transverse Fabry-P\'erot modes).\cite{Ihn:2009} In all of these cases constructive interference occurs at a series of resonant wavelengths and the resulting localization of the charge carriers manifests in reduced conductance. The charge carrier wavelength can be manipulated by bias voltage or gate voltage, such that quantum interference can be directly observed in the conductance as a function of gate voltage and bias voltage.\\
Fig. \ref{Tdep} shows the conductance $G$ at different temperatures from 30 mK to 28 K in device B as a function of gate voltage $V_\text{g}$ with a small Coulomb blockade region at low gate voltage and quantum interference features for a more open channel. The amplitude of the quantum interference features is decreasing as the temperature is increasing until the $G$-$V_\text{g}$ trace is nearly smooth at 28 K. In a single conducting channel without averaging over independently fluctuating segments Universal conductance fluctuations are expected to reach amplitudes $\alpha e^2/h$ with $e^2/h=38.7\mu\text{S}$ and $\alpha$ on the order of 1 depending on device geometry.\cite{Ihn:2009,Yao:2012} In our data the largest fluctuation produces only $\alpha=0.12$ which could be related to averaging over multiple channels or an unaccounted series resistance.\\
The inset of Fig. \ref{Tdep} shows the root-mean-square of the $G$-$V_\text{g}$ traces $\Delta G$ as a function of temperature after removing the background (the procedure for obtaining $\Delta G$ is described in the Supplementary Information). Below 1.9 K $\Delta G$ does not depend on temperature, as predicted for the transport regime where the coherence length $l_\phi$ is longer than the nanowire, such that there is no averaging over independently fluctuating segments of the nanowire.\cite{Yao:2012} 

For temperatures above 1.9 K the conductance fluctuations follow a power law $\Delta G \propto T^{\gamma}$ (see Supplementary Information)\cite{Ihn:2009,Echternach:1993}. We can calculate the expected value for $\gamma$ assuming that the coherence length is proportional to $T^{-1/3}$ and the dominant dephasing mechanism is 1D electron-electron interaction\cite{Echternach:1993}. If the thermal broadening of the electron energy distribution, that is parameterized by the thermal length $l_\text{T}$, is larger (smaller) than the coherence length we expect a power $\gamma$ of -2/3 (-1/2). We fit the data for temperatures above 1.9 K with a power law using the power $\gamma$ as a free parameter and extract a value of $\gamma=-0.67\pm0.04$ in excellent agreement with the case $l_\text{T}<l_\phi$.\\

Figure \ref{waterfallandstab}a shows the conductance $G$ as a function of bias voltage $V_\text{b}$ for gate voltages from 0.97 V to 7.5 V. The red lines mark the traces at 2, 3, 4, 5, 6 and 7 V in gate voltage for clarity. All conductance traces show a dip centered around zero bias that is a signature of strong electron-electron interactions and typical for one-dimensional systems.\cite{Tilke:2003,Lu:2005,Mirza:2017} In case there is only little change in the conductance as a function of gate voltage the lines in Fig. \ref{waterfallandstab}a are close together and the plot appears dark. Away from the Coulomb-blockade region these darker features have been related to quantum interference and are expected to show a characteristic pattern of alternating zero-bias and non-zero bias features.\cite{VanWeperen:2013} The zero-bias features correspond to a resonance with the Fermi-energy of the charge carriers, while the non-zero bias features correspond to the point where the bias window encompasses the energies of a neighboring resonance. In Fig. \ref{waterfallandstab}a the resulting pattern is visible between 5 and 6 V in gate voltage as well as between 6 and 7 V but other non-zero bias features are faint or not visible.\\
To visualise the quantum interference features in a different way we show the transconductance  $dG/dV_\text{g}$ as a function of gate voltage $V_\text{g}$ and bias voltage $V_\text{b}$ in Fig. \ref{waterfallandstab}b. For lower gate voltages ($V_\text{g} \lesssim 1.7$ V) we observe a large Coulomb gap and some Coulomb-blockade features that correspond to multiple charge islands in series (see Supplementary Information). At higher gate voltages, when the overall conductance increases (see Fig. \ref{waterfallandstab}a), quantum interference results in diamond shaped patterns as a function of gate- and bias voltage and we find the corresponding features from Fig. \ref{waterfallandstab}a.\\

In the remainder of this letter we will analyze the characteristic energy spacings of the conductance fluctuation pattern to infer the microscopic origin of the quantum interference and find the elastic mean free path $l_\text{e}$. To determine different transport regimes, we employ the definitions introduced by Beenakker and van Houten. In the diffusive transport regime both the wire diameter $W$ and length $L$ are much larger than the elastic mean free path. However elastic impurity scattering does not destroy phase coherence, such that effects of quantum interference can still modify the conductivity of the disordered conductor. Transport is considered to be ballistic when the dimensions of the wire are reduced below the mean free path. The intermediate quasi-ballistic regime is characterized by $W<l_e<L$, meaning boundary scattering and internal impurity scattering are of equal importance. At low temperatures the phase coherence length $l_{\phi}$ can extend over a large part of the wire and exceed $L$, resulting in conductance fluctuations when the transport is quasi-ballistic or even diffusive. By comparing the elastic mean free path to the sample dimensions we will determine whether the quantum interference is dominated by boundary scattering or impurity scattering.\\

We analyze the energy spacings of the conductance fluctuations in two different devices with length 150 nm and diameter 8 nm: device B (from the last sections) with activated P doping density of $4\times10^{19}\,\text{cm}^{-3}$ and device C with a higher activated P doping density of $2\times10^{20}\,\text{cm}^{-3}$. Fig.~\ref{FFTauto}a shows the transconductance $dG/dV_{g}$ as a function of gate voltage $V_\text{g}$ and bias voltage $V_\text{b}$ for device B at 13 mK and \ref{FFTauto}b shows the transconductance for device C at 4 K.

To extract the energy scales we calculate the autocorrelation function of both data sets and take the Fourier transform to find the dominant voltage spacings between the fluctuations.\cite{Oksanen:2014} The result for device B is plotted in Fig.~\ref{FFTauto}c and shows four symmetric hotspots (circled in white) corresponding to a gate voltage spacing $\Delta V_\text{g}=0.6\pm0.1\,\text{V}$ and bias spacing $\Delta V_\text{b}=30\pm26\,\text{mV}$. For device C the same analysis in Fig.~\ref{FFTauto}d shows hotspots corresponding to $\Delta V_\text{g}=1.5\pm0.2\,\text{V}$ and $\Delta V_\text{b}=50\pm30\,\text{mV}$. The extracted gate voltage spacings from the two samples reflect the different energy spacings in the corresponding transconductance plots Fig.~\ref{FFTauto}a,b. The periodicity in bias voltage does not give an accurate picture of the energy spacings, since there are hardly multiple features on the bias axis in Fig. \ref{waterfallandstab} and instead is likely related to artifacts from the limited bias range of the data.\\
We convert the extracted gate-voltage spacings into the characteristic length scales of the quantum interference to find the elastic mean free paths $l_\text{e}$ and compare them to the length scales in the device to identify the origin of the effect. In a simple particle-in-a-box picture we can associate oscillations on the gate voltage axis to the characteristic length scale of the quantum interference, and thus the elastic mean free path, using:\cite{Biercuk:2005}
\begin{equation}
l_\text{e}=\frac{4e}{c_\text{g}\Delta V_\text{g}}\label{gatetolength}
\end{equation}
Here $e$ is the elementary charge, $\Delta V_\text{g}$ is the distance in gate voltage between two features and $c_\text{g}$ is the capacitance to the gate per unit length.
The capacitance per unit length $c_\text{g}=(1.1\pm0.2)\times10^{-10}\,\text{F/m}$ can be taken from the analysis of Fig. \ref{CB} where we observed capacitive coupling to the entire channel of the geometrically identical device A.\\
Following Eq. \ref{gatetolength} we convert the dominant gate voltage spacing in device B to $l_\text{e}=10\pm2\,\text{nm}$ and in device C to $l_\text{e}=4\pm2\,\text{nm}$. In device C this yields $l_\text{e}< W< L$ (as illustrated in Fig. \ref{gatetolength}f) at 4 K and a mean free path that is close to the average spacing of unactivated dopants (1-2 nm) that might be larger in the bulk of the nanowire due to surface segregation.\cite{Bjork:2009} This result therefore agrees with previous mobility measurement that indicated dominant scattering from neutral impurities.\cite{Mirza:2014} Considering the prominence of surfaces in such small diameter nanowires this is a surprising result that underpins the effectiveness of the forming gas anneal in reducing the number of interface trap states.\\ 
In device B we find $W \lesssim l_\text{e}  < L < l_\phi$ at 13 mK such that scattering from the radial constraints and the neutral impurity scattering contribute equally to the observed quantum interference as schematically depicted in Fig. \ref{gatetolength}e. In these devices the average distance between unactivated dopants is around 5~nm, but surface segregation could again increase this spacing in the bulk of the nanowire.\cite{Mirza:2014} By reducing the doping concentration by 1 order of magnitude between device C and B we are able to make the transition between the diffusive transport regime, where the dominant scattering source are neutral impurities, to the quasi-ballistic transport regime, where the contribution of boundary scattering and impurity scattering are equal. This provides us with a viable pathway towards a fully ballistic silicon nanowire, either by reducing the channel length to less than 10~nm, or by reducing the doping concentration.\\ 

In this letter we measure transport in three highly P doped Si nanowires with diameter $8\pm0.5$ nm, length 150 nm and a wrapped around gate. One nanowire shows regular Coulomb blockade over a gate range of 6 V that we attribute to strong electrostatic confinement of electrons into an island that extends along the length of the nanowire and can therefore be used to extract the intrinsic gate capacitance of the nanowire geometry.\\ 
In two other devices with larger conductance, we observe quantum interference features that can either originate from random potential fluctuations along the nanowire or transverse modes. We extract an elastic mean free path of $4\pm2$ nm in a device doped Si:P $2\times10^{20}\,\text{cm}^{-3}$, which is close to the average spacing of the unactivated dopants. In the lower doped device with $4\times10^{19}\,\text{cm}^{-3}$ the elastic mean free path is $10\pm2\,\text{nm}$, which is larger than the diameter and therefore indicates quasi-ballistic transport. Temperature dependent measurements in this device show the expected behavior for Universal Conductance Fluctuations and suggest coherence lengths larger than the nanowire length at low temperatures as well as dephasing due to electron-electron interactions at temperatures above 1.9 K. In agreement with previous mobility measurements\cite{Mirza:2014} these results suggest that the dominant scattering process at low temperatures is impurity scattering rather than scattering due to surface effects that are often dominant in nanowire devices.\\
We have demonstrated a junctionless, 1D quasi ballistic transistor that could be optimized to reach the ballistic limit for shorter channel lengths and lower doping density. The top-down fabrication of our silicon devices is one of the key requirements for CMOS integration making our devices technologically relevant for applications in low power or cryogenic CMOS as well as a platform for quantum electronic devices including charge pumps and charge sensors.\cite{Schupp:2017} \\

\section*{Acknowledgement}
The work was funded by the U.K. EPSRC (grants
EP/H024107/1, EP/N017188/1 and EP/N003225/1) and DSTL contract
1415NatPhD\_59. The authors would like to thank Pascal Gehring for helpful discussions and the staff of the
James Watt Nanofabrication Centre for help and support
in undertaking the research.

\bibliography{libraryDoug.bib}
\bibliographystyle{achemso.sty}
\end{document}